# Enhancing the Resolution Limits of Spectral Interferometric Measurements with Swept-Wavelength Interrogation by Means of a Reference Interferometer

NIKOLAI USHAKOV,[1,*] ALEXANDER MARKVART,[1] LEONID LIOKUMOVICH[1]

[1] *Department of Radiophysics, St. Petersburg State Polytechnical University, Polytechnicheskaya, 29, 195251, St. Petersburg, Russia*
*Corresponding author: n.ushakoff@spbstu.ru

**An approach for compensating the influence of the interrogator noises on the readings of interferometric sensors, interrogated by means of spectral interferometry with wavelength tuning is proposed. Both theoretical analysis and a proof-of-principle experiment were performed for the example of extrinsic Fabry-Perot interferometers (EFPI). Two schemes, comprising of a signal and a reference interferometers, switched in different optical channels of the interrogating unit, were proposed. The approach is based on the fact that the fluctuations of some of the interrogator parameters produce correlated fluctuations of the reference and signal interferometers' optical path differences' (OPD) measured values. The fluctuations of the reference interferometer's measured OPD can be subtracted from the measured OPD of the signal interferometer. The fluctuations of different parameters of the interrogator are considered, the correlation properties of the produced noises of the measured OPD values are demonstrated. The first scheme contains two interferometers with similar parameters and enabled a three-fold resolution improvement in the performed experiments, when the difference of the interferometers OPDs was varied within about 10 nm. The second scheme contains two interferometers with OPDs difference such that all interrogator fluctuations, except for the dominating one produce uncorrelated OPD errors. With the second scheme a two-fold resolution improvement was experimentally demonstrated, when the interferometers' OPDs difference was varied within more than 1 μm. The proposed approach can be used for improving the resolution of interferometric sensors with relatively large OPDs (greater than 200-300 μm), which can be advantageous for remote materials and surface inspection. The other potential application is the use of relatively simple and chip interrogators with poor wavelength scale repeatability for high-precision measurements.**

http://dx.doi.org/10.1364/AO.99.099999

## 1. Introduction

During the last two decades a great progress in manufacture and implementation of the fiber optic sensors based on the extrinsic Fabry-Perot interferometers (EFPI) [1] has been achieved by the academic institutions and commercial companies. Such sensors demonstrate a great dynamic measurement range and a high resolution [2,3]. Sensors of a great variety of physical quantities have been designed and implemented. The most commonly used EFPI optical path difference (OPD) demodulation approaches are the white-light interferometry, using a scanning readout interferometer [4,5] and spectral interferometry, in which the measurement and further processing of the interrogated interferometer spectral transfer function is used to measure the interferometer OPD [6–9]. The best attained OPD resolutions are about 15-40 picometers [2,10,11], estimated as doubled standard deviation (STD) of the measured OPD fluctuations $\sigma_{Lr}$. All such high resolutions were attained in case of short-cavity sensors with OPD value less than 100 μm. An analysis of the noise sources and the light propagation inside the EFPI cavity was made in a paper [10], dedicated to resolution limits analysis for single EFPI sensors.

Throughout this paper we consider the case of registering the spectrum of the light reflected from the sensor, which is the most common way of interrogation. The spectral function of a low-finesse

Fabry-Perot interferometer in this case is expressed as $S_{FP}(L,\lambda) = S_0(L,\lambda) + S_M \cdot S(L,\lambda)$, where [10]:

$$S(L, \lambda) = \cos[4\pi nL/\lambda + \gamma_R(L, \lambda)], \quad (1)$$

$$S_M = 2\sqrt{R_1 R_2 \cdot \eta(L)}, \quad \eta(L) = \frac{(\pi n w_0^2)^2}{L^2\lambda^2 + (\pi n w_0^2)^2}, \quad (2)$$

$$\gamma_R(L,\lambda) = \psi_R + \varphi_R = \operatorname{atan}(4L^3/z_R^3 + 3L/z_R) + \varphi_R, \quad (3)$$

and an approximation of Gaussian profile was applied to the fiber mode and the beam inside the interferometer. $\eta(L)$ is a coupling coefficient of a light beam, irradiated by a fiber mode and travelled a distance $2L$ back to a fiber mode [10]; $L$ is the interferometer OPD; $w_0$ is an effective radius of a mode at the output of the first fiber; $\lambda$ is the light wavelength; $n$ is the refractive index of the media between the mirrors; $z_R = \pi n w_0^2/\lambda$ – Rayleigh length of the Gaussian beam; the argument additive $\gamma_R(L,\lambda)$ contains a phase shift $\psi_R$, induced by the light diffraction inside the cavity, and a phase $\varphi_R$, induced by the mirrors (typically for dielectric mirrors $\varphi_R=\pi$). The equations above are valid for the case of parallel mirrors.

One can distinguish several classes of approaches for obtaining the OPD estimate $L_r$ from the registered spectrum. Among them are the so-called "frequency-estimating" [7,12], based on the analysis of the oscillation frequency of $S'(\lambda)$ fringes; and the so-called "phase-estimating" [12–14], evaluating the argument of the $S'(\lambda)$ fringe oscillations; and the approximation-based [3,6], approximating the $S'(\lambda)$ with analytical expression (1) by means of a least-squares fitting. The latter two approaches have proved their higher efficiency via various experimental and numerical studies. In the current paper the approximation-based approach [3] is utilized.

As a result of various noise mechanisms influences [10–12], the estimated quantity $L_r$ will differ from the real interferometer OPD $L_0$. Due to stochastic nature of the noise mechanisms, the estimated OPD value will fluctuate from measurement to measurement, resulting in a finite measurement resolution, determined by $\sigma_{Lr}$ (STD of the $L_r$ fluctuations).

An idea of compensating hardware noises in interferometric sensing systems had been proposed long ago, mainly dedicated to reducing the laser frequency noise influence [15]. In a paper [16] the idea of noise reduction with a compensating interferometer was realized for extrinsic Fabry-Perot sensors, interrogated by means of the spectral interferometry. As a result, the sensor resolution was tenfold improved (from 700 nm to 70 nm). However, this approach is mostly dedicated to the elimination of the environmental influence and isn't applicable for suppression of the hardware-induced noises impact. In the current paper an approach, able to suppress the influence of the interrogating unit noises on the sensor resolution, is proposed.

## 2. Noise mechanisms

An extensive study of single EFPI displacement sensors resolution limits with wavelength-scanning interrogation was done in [10]. It was shown that the main noise sources are:

1. Absolute wavelength scale shift $\Delta\lambda_0$, determined by fluctuations of the triggering of the scanning start, $\sigma_{\Delta\lambda}=\operatorname{stdev}\{\Delta\lambda_0\}$.
2. Jitter of the wavelength points $\delta\lambda_i$, caused by the fluctuations of the signal sampling moments, $\sigma_{\delta\lambda}=\operatorname{stdev}\{\delta\lambda\}$.
3. Additive noises $\delta s_i$, produced by the photo registering units, by the light source intensity noises, etc. $\sigma_S=\operatorname{stdev}\{\delta s\}$.

These mechanisms will result in distortion of the registered interferometer spectral function $S'(\lambda)$. Therefore, the spectrum approximation procedure gives an erroneous result, denoted throughout the paper as $L_r$. When considering a vector of consequently measured OPD values, its standard deviation $\sigma_{Lr}$ can be calculated. Generally, $\sigma_{Lr}$ is used as a quantitative characteristic of a sensor resolution, which is approximated as $2 \cdot \sigma_{Lr}$.

The first mechanism provides the shift of the measured interferometer spectrum, inherently shifting the displacement sensor readings as follows

$$\delta L \approx -\Delta\lambda_0 \cdot L_0/\lambda_0, \quad (4)$$

where $\lambda_0$ is the central wavelength of the interval, on which the interferometer spectral function $S'(\lambda)$ is measured.

The jitter of spectral points during interrogation produces the distortion of the measured spectral function $S'(\lambda)$. This distortion can be interpreted as an additive noise with some variance. The resulting signal-to-noise ratio $\mathrm{SNR}_{JIT}$ can easily be estimated by simple trigonometric derivations [10]:

$$\mathrm{SNR}_{JIT} = \frac{2\lambda_0^4}{[8\pi n L_0 \sigma_{\delta\lambda}(f)]^2}, \quad (5)$$

where frequency dependence of the wavelength jitter $\sigma_{\delta\lambda}(f)$ was introduced. In turn, the interferometer OPD $L_0$ can be interpreted as a quantity, related to the frequency of the spectral transfer function oscillations. Since the signal $S'(\lambda)$ is registered by means of the wavelength sweeping, it can be treated as a temporal quasi-harmonic signal [17], oscillating with frequency, close to

$$f_S = 2k_\lambda n L_0/\lambda_0^2, \quad (6)$$

where $k_\lambda$ is the wavelength tuning speed of the laser.

Considering the third mechanism, one has to take into account that generally the noise variance can depend on the mean optical power $P_I$, incident to the photodetector (shot noise level and laser intensity noise influence are strongly related to the mean power level). The dependency can be adequately approximated by a power function as follows

$$\sigma_s = P_I \cdot \mathrm{RIN}(f) + \mathrm{NEP}, \quad (7)$$

the second variant of (7) is given with taking into account that in conventional optical sensors interrogators, like the one used in the current work, thermal and electronic photodetector noises (determined by NEP) dominate the shot noise, RIN is parameter, characterizing laser intensity noises, and can be frequency-dependent, as well as the wavelength jitter. In case of EFPI, the noise, given by eq. (7), produces the SNR of the measured spectral function $S'(\lambda)$, written as follows

$$\mathrm{SNR}_s = \frac{S_m^2/2}{\sigma_s^2} = \frac{2P_0^2 R_1 R_2^*}{P_0^2(R_1+R_2^*)^2 \mathrm{RIN}^2(f) + \mathrm{NEP}^2}, \quad (8)$$

where $P_0$ is the optical power irradiated by the light source; $R_2^* = R_2 \cdot \eta$ is effective mirror reflectivity, taking into account the light losses caused by the divergence of a non-guided beam inside the cavity.

The influence of the additive noises on the STD of the measured OPD values $L_r$ can be found either by numeric simulation, or analytically using the Cramer-Rao bound [18,19]. It can be shown that for the approximation approach [3] the corresponding relation can be written as

$$\sigma_{Lr}(\mathrm{SNR}) = C \cdot \mathrm{SNR}^{-1/2}, \quad (9)$$

where $C$ is within the interval $(9 \div 11) \cdot 10^{-4}\,\mu\mathrm{m}^{-1}$ for different parameters of the OPD measurement approach [3].

Finally, the expression for the OPD standard deviation can be obtained by combining expressions (4) and (9) and taking into account the variance summation rule:

$$\sigma_{L_r} = \left\{ \left(\frac{L}{\lambda_0}\right)^2 \cdot \sigma_{\Delta\lambda}^2 + C^2 \left[ \frac{P_0^2(R_1+R_2\eta(L))^2 \mathrm{RIN}^2(f) + \mathrm{NEP}^2}{P_0^2 R_1 R_2 \eta(L)} + \frac{[8\pi n L \sigma_{\delta\lambda}(f)]^2}{2\lambda_0^4} \right] \right\}^{1/2}. \quad (10)$$

The equation (10) can be used in order to consider partial impacts of different mechanisms into the overall $\sigma_{Lr}$ value. In fig. 1 a comparison of noise-induced and scale shift-induced measured OPD fluctuations for two EFPI configurations ($R_1=R_2=3.5\%$ and $R_1=3.5\%$, $R_2\approx 90\%$) is presented. A comparison of analytical predictions with experimental data, obtained in [10] is also presented, demonstrating good correspondence, and hence, the adequacy of the analytical results.

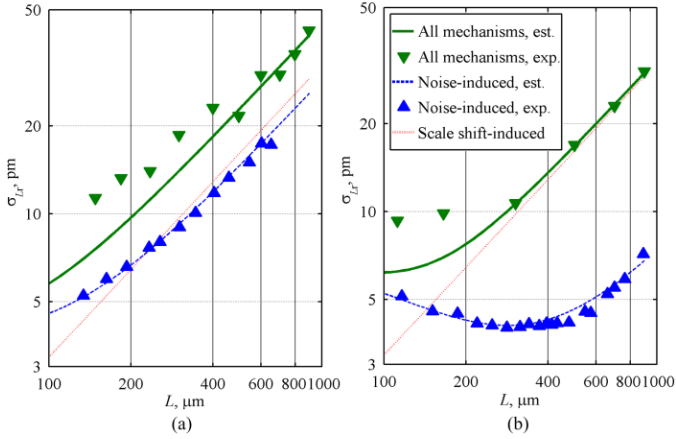

Fig. 1. Comparison of noisy mechanisms for EFPIs with $R_1=R_2=3.5\%$ (a) and $R_1=3.5\%$, $R_2\approx 90\%$ (b). Experimental data from [10].

Such simple manner of introducing the frequency dependence of noises as in (5) and (7) is possible since the signal $S'(\lambda)$ is quasi-harmonic and when it is processed by the approximation procedure, the error of determining the interferometer OPD $L_r$ is stipulated only by those noise's spectral components, which frequencies coincide with the $S'(\lambda)$ oscillation frequency, given by (6).

## 3. Theoretical analysis of noise compensation

Throughout this paper the following noise compensation scheme will be assumed: two interferometers are used for the measurement, one of them is exposed to the target perturbation (which is the actual measured quantity) and will be referred to as sensing or signal interferometer, and the second one is isolated from any environmental changes and will be referred to as reference interferometer. The interferometers' OPDs will be denoted $L_S$ and $L_R$, respectively. Both these interferometers will be assumed to be interrogated by the same tunable laser, while their spectral functions will be registered by independent (although, similar) photodetectors.

In order to compensate the parasitic OPD fluctuations, one has to find the OPDs of both the signal and reference interferometers (the measured OPD values will be denoted as $L_{rS}$ and $L_{rR}$), after that subtract the deviations of the reference interferometer OPD with respect to its initial value from the sensing interferometer OPD values:

$$L_{rSC}(t) = L_{rS}(t) - [L_{rR}(t) - L_{rR}(0)]. \quad (11)$$

Considering the efficiency of the noise compensation, one needs to determine the likelihood of the two interferometers' measured OPDs fluctuations, which can be done by means of a correlation coefficient. Therefore, in order to study the ability of noise cancellation, one needs to study the behavior of the correlation coefficient $C_{BF}$ between the registered OPD values of reference and sensing interferometers with respect to the mechanisms, mentioned in section 2. As can be concluded from the eq. (10), the closer are the interferometers' parameters, the higher will be the correlation of $L_{rS}$ and $L_{rR}$ fluctuations. Therefore, at first, let us consider the case of similar signal and reference interferometers.

### A. OPD noises statistics

As can be seen from expression (4), the absolute shift of the wavelength scale produces an error, proportional to the OPD value; therefore, its influence will be correlated for the signal and reference interferometers even with arbitrarily different OPDs.

Considering the impact of individual spectral points jitter, one has to take into account that it is indirect in nature: the jitter itself produces a distortion of the measured spectral function, which can be interpreted as additive noises. These noises, in turn, affect the accuracy of the approximation procedure. However, in most interrogators, these noises are produced by different unsynchronized triggering systems of photodetectors and analog to digital converters, used to digitize the spectral functions in different optical channels of the interrogator, and, therefore, are uncorrelated, producing uncorrelated OPD measurement errors.

Considering the additive noises (the third mechanism), one needs to take into account their two main sources – laser intensity noises and the photodetector noises. Laser intensity noises, being the same for the measurement and reference interferometers, will produce identical errors in case of close OPD values $L_S \approx L_R$ and $|L_S - L_R| = \lambda_0/2$. In case of $|L_S - L_R| = \lambda_0/4$ the influence of laser intensity noises will be inverse for signal and reference interferometers and will produce anti-correlation $C_{BF}(\lambda_0/4) = -1$. For arbitrarily different OPDs the OPD fluctuations will be uncorrelated, $C_{BF} = 0$. Dependency $C_{BF}(L_S - L_R)$ for a particular setup will depend on the relation of the above mentioned noise sources. Photodetector noises, which are produced by different devices, and hence, are uncorrelated, will produce uncorrelated errors.

In order to study the exact dependency of the signal and reference interferometers' OPD noises correlation $C_{BF}$ on the OPD difference, a numeric simulation has been performed. Two spectral functions $S(L_{S,R}, \lambda)$ were calculated according to (1), the difference $\delta L = L_S - L_R$ was varied within the interval $0 \div 0.8$ μm, the values $L_{S,R}$ themselves were around 200 μm. The rest parameters were chosen close to the ones in a later performed experiment (see section 4): wavelength scanning range $1510 \div 1590$ nm, interval between spectral points $\Delta = 4$ pm, resulting in $M = 20001$ points in spectrum; wavelength jitter stdev $\sigma_{\delta\lambda} = 1$ pm; scale shift stdev $\sigma_{\Delta\lambda} \approx 0.05$ pm; laser output optical power $P_0 \approx 0.06$ mW; RIN in the full frequency band -50 dB (-110 dB/Hz for ~ 1 MHz photodetector frequency band); photodetector NEP 80 pW in the full frequency band ($8\cdot 10^{-13}$ W/√Hz for ~ 1 MHz photodetector frequency band). Modeling was performed for two interferometers schemes: with equal mirrors $R_1=R_2=3.5\%$ and with second opaque mirror $R_1=3.5\%$, $R_2\approx 90\%$. For all OPD combinations $L_S$, $L_R$ an ensemble of $N=100$ realizations of noises were calculated for better statistical validity of the simulation results. Two cases were considered: an individual influence of laser intensity noises and joint influence of all noise sources. The correlation coefficients $C_{BF}$ of the resulting vectors of the OPD fluctuations were calculated.

As was observed during the modeling, when only the laser intensity noises (modelled as additive noises equal for the two spectral functions $S(L_S, \lambda)$ and $S(L_R, \lambda)$) were presented, the measured OPDs correlation coefficient $C_{BF}$ was equal to the correlation coefficient between the spectral functions $S(L_S, \lambda)$ and $S(L_R, \lambda)$.

The dependencies $C_{BF}(\delta L)$, modelled in case of simultaneous influence of all noise sources are shown in fig. 2 as solid lines.

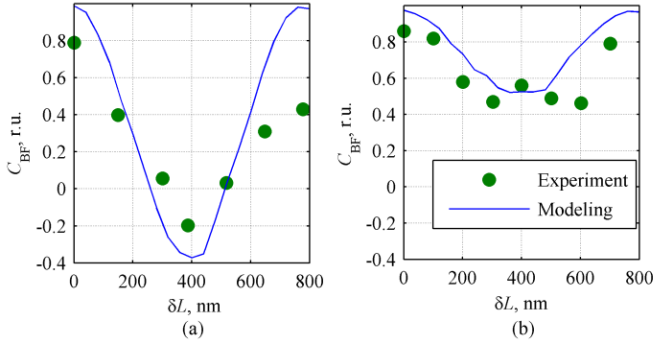

Fig. 2. Dependencies of the measured OPD fluctuations correlation coefficient on the OPD difference for all noise mechanisms: modelling (line) and experimental (points) for interferometers with $R_1=R_2=3.5\%$ (a) and $R_1=3.5\%$, $R_2\approx90\%$ (b).

### B. Noise compensation efficiency

The value of the correlation coefficient itself doesn't reflect the particular reduction of the $L_{rS}$ fluctuations that can be attained by means of the noise compensation, performed according to the eq. (11). Quantitatively this reduction can be characterized by the ratio of STDs of the registered OPDs without and with compensation:

$$E = \sigma_{LrS}/\sigma_{LrSC}, \quad (12)$$

where $\sigma_{LrS}$ and $\sigma_{LrSC}$ are STDs of $L_{rS}$ and $L_{rSC}$, respectively, $E$ will further be entitled as compensation effort. Mathematical derivations, describing the relation of the measured OPDs correlation coefficient $C_{BF}$ and the compensation effort, are presented in Appendix A, the dependency $E(C)$, calculated according to the eq. (A4) is shown in fig. 3.

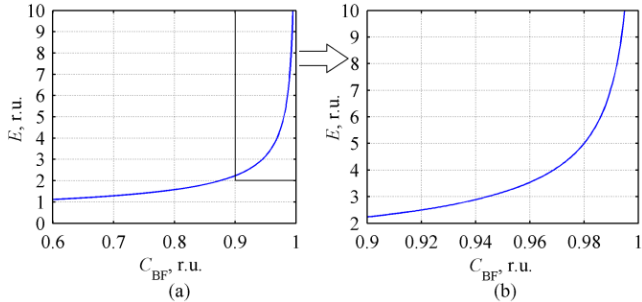

Fig. 3. Dependency of the compensation effort on the correlation coefficient of the registered OPDs fluctuations $L_{rS}$ and $L_{rR}$. The plot in (b) is an enlargement of a fragment highlighted in (a).

As can be concluded from this dependency, for considerable noise compensation (at least two-fold reduction of the noise STD), the correlation coefficient $C_{BF}$ must be at least 0.87. As can be seen from the modelled dependencies $C_{BF}(\delta L)$, illustrated in fig. 2, in case of equal mirrors $R_1=R_2=3.5\%$ this condition is fulfilled for OPDs differences $\delta L<50$ nm and in case of $R_1=3.5\%$, $R_2\approx90\%$ – for $\delta L<100$ nm.

In an ideal case of equally adjusted interferometers OPDs, the influences of the both laser intensity noise and wavelength scale shift on the measured OPDs will be totally compensated, with no affection on the fluctuations of the measured $L_{rSC}$ values. For estimation of the resolutions, attainable in this case, let us substitute into (10) the following parameters: $\sigma_{\Delta\lambda}=0$, $\sigma_{\delta\lambda}=\sqrt{2}\cdot\sigma_{\delta\lambda}$, RIN=0, NEP=$\sqrt{2}\cdot$NEP (two photodetectors will produce independent additive noises and wavelength points jitter of the same level, the impact on the $\sigma_{LrSC}$ will therefore be stipulated by doubled variance of a single detector influence). In such a manner, the eq. (10) is modified to the following form:

$$\sigma_{L_rSC} = \sqrt{2}\sigma_{L_r}\Big|_{NEP,\delta\lambda} = \frac{\sqrt{2}C}{\sqrt{SNR}} =$$

$$C \cdot \left( \frac{NEP^2}{R_1 R_2^* \cdot P_0^2} + \frac{[8\pi n L_0 \sigma_{\delta\lambda}(f)]^2}{2\lambda_0^4} \right)^{1/2}. \quad (13)$$

However, in practical interrogating unit the wavelength scale shift in different optical channels may be not absolutely the same, therefore, reducing the efficiency of the noise compensation.

### C. Non-matched interferometers

Even though the potentially attainable compensation effort in scheme with similar interferometers is considerable, the practical applicability of such resolution enhancement approach is limited due to strong requirement on the interferometers' OPDs matching accuracy, which, for an acceptable compensation effort must be better than ~10 nm in the implemented experimental setup, therefore, limiting the dynamic range of the measurements to somewhat less than $10^4$. This limitation can be overcome by the use of adjustable reference interferometer with variable OPD and a feedback. However, this will complicate the practical system and degrade the possible measurement speed. Nevertheless, as was observed during the numeric experiments (see fig. 2), the decrease of the correlation $C_{BF}$ for OPD offset values $\delta L \sim \lambda_0/4$ is mainly due to the laser intensity noise influence, causing anti-correlated fluctuations of $L_{rS}$ and $L_{rR}$. In case of the further increase of $\delta L$, the correlation of $L_{rS}$ and $L_{rR}$ fluctuations will demonstrate decaying oscillating character with quasi-period of $\lambda_0/2$. Therefore, another way to improve the dynamic range of the measurements is to eliminate the influence of the laser intensity noise, or, at least, to make it independent for the reference and the measurement interferometers.

As was discussed in section 3.1, the correlation coefficient $C_{BF}$ of measured OPD fluctuations in case of equal additive noises is equal to correlation coefficient of the interferometers' spectral functions $C_{SF}$. The dependency of $C_{SF}$ on the interferometers OPDs difference $\delta L$ is illustrated in fig. 4. It can be seen that $C_{SF}(\delta L)$ dependency has a decreasing oscillating form with envelope determined by $sinc(\delta L/\Delta L)$ function, where $\Delta L$ is an equivalent resolution in the interferometer OPD domain, expressed as

$$\Delta L \approx \lambda_0^2/2n\Lambda, \quad (14)$$

where $\Lambda$ is the width of the wavelength interval, on which the spectral function is measured.

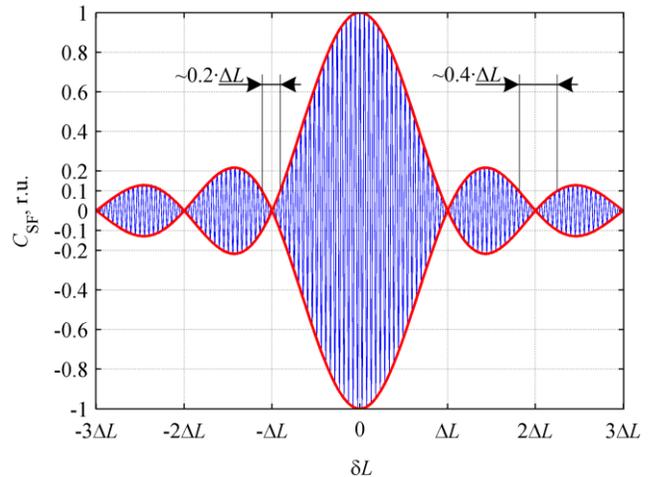

Fig. 4. Dependency of correlation coefficient of the reference and the signal interferometers' spectral functions on their OPDs difference. The thick line highlights the envelope of the dependency.

As can be seen from the fig. 4, in case of δL close to either ΔL or 2ΔL, corresponding to zeros of sinc(δL/ΔL) function, the correlation coefficient of $S(L_S,\lambda)$ and $S(L_R,\lambda)$ spectral functions is close to zero, therefore, as can be predicted from the results of numeric modeling, the correlation coefficient of the OPD fluctuations will be close to zero as well. In fig. 4 the δL intervals, on which the absolute value of the correlation $C_{SF}$ is below 0.1 are highlighted and their widths are shown. For the parameters of the experimental setup, used in the current work (see section 4), ΔL≈15µm, therefore, the widths of the intervals for possible OPD difference variation is greater than 1 µm, increasing the attainable measurement dynamic range at about two orders. Moreover, for total subtraction of the OPD error, induced by the wavelength scale shift, the eq. (11) must be modified to

$$L_{SC}(t) = L_S(t) - L_S(0)/L_R(0) \cdot [L_R(t) - L_R(0)]. \quad (15)$$

However, the best attainable resolution will degrade comparing to the one given by eq. (13) due to the influence of the laser intensity noise. The corresponding expression can be obtained from the eq. (10) by substituting $\sigma_{\Delta\lambda}=0$, $\sigma_{\delta\lambda}=\sqrt{2}\cdot\sigma_{\delta\lambda}$, RIN=$\sqrt{2}\cdot$RIN, NEP=$\sqrt{2}\cdot$NEP, and is expressed as follows

$$\sigma_{L_rCN} = \sqrt{2}C\left[\frac{P_0^2(R_1+R_2\eta(L_S))^2 \text{RIN}^2(f)+\text{NEP}^2}{P_0^2 R_1 R_2 \eta(L_S)} + \frac{[8\pi n L_0 \sigma_{\delta\lambda}(f)]^2}{2\lambda_0^4}\right]^{1/2}. \quad (16)$$

An assumption $\delta L \ll L_S$ was used during the derivation of the eq. (16), which is valid in our case: in the experimental setup, used in the current work, the OPD values $L_{S,R}$>200 µm are required for the wavelength scale shift influence be dominating the other mechanisms (see fig. 1), therefore, the condition $L_{S,R} \gg \Delta L \sim 10$ µm will be fulfilled anyhow.

### D. Additional noise sources

In the current paper in calculations according to the developed analytical model all noises will be assumed white. However, as can be concluded from eqs. (10), (13) and (16), in case of nonuniform noise spectra the dependencies of measured OPDs' fluctuations' STD on the interferometers OPD values will be different from those in case of uniform noise spectra. This may be one of the reasons for discrepancies of the experimentally measured $\sigma_{Lr}$ values and values, estimated analytically.

Another possible source of additional noises is parasitic interference components, stipulated by alternative light propagating paths inside the interrogator (reflections on fiber connectors, higher-order mode propagation). In fig. 5 examples of such components (can be observed as peaks at abscissa points around 495.1 and 3.213e+04) are shown for an example of experimentally measured spectral function. Under the influence of the wavelength points jitter these interference components will produce additive noises with level

$$\sigma_{nPI} = S_{PI}/\text{SNR}_{PI}^{1/2} = \frac{8\pi L_{PI}\sigma_{\delta\lambda}(f)\cdot S_{PI}[R_1+R_2\eta(L_0)]}{\sqrt{2}\cdot\lambda_0^2}, \quad (17)$$

where $L_{PI}$ is OPD corresponding to the parasitic interference component, $S_{PI}$ is the amplitude of the component itself.

In turn, these parasitic interference components in different optical channels will probably be stipulated by different optical elements (this will be the case if the light propagates through these parasitic paths after the interferometers) and therefore, produce uncorrelated fluctuations of measured OPDs. Such situation is in some sense analogous to serial multiplexing of interferometers, considered in [20]. Therefore, amplitudes of these parasitic interference components will be proportional to the mean level of the interferometer spectral function $S'(\lambda)$.

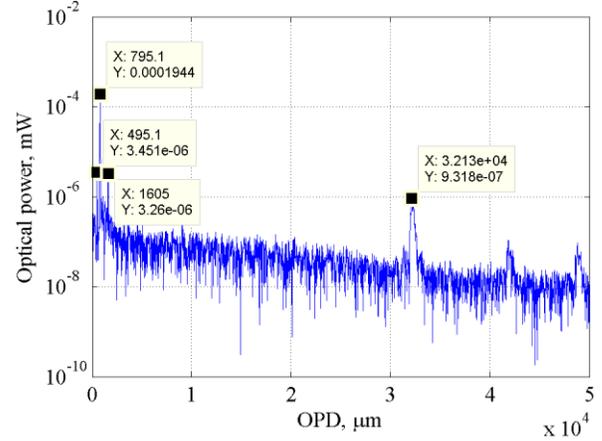

Fig. 5. Fourier transform of a measured spectral function of EFPI with $R_1=R_2$=3.5%, OPD≈800 µm.

Estimating the influence of these parasitic interference components on the noise compensation, an additional term must be added in eqs. (10), (13) and (16). Assuming that the levels of these components are equal in the two optical channels, the equation for $\sigma_{Lr}$ with taking into account the parasitic components' influence, can be found as

$$\sigma_{L_rPI} = \left[\sigma_{Lr}^2 + \frac{4C^2\sigma_{nPI}^2}{P_0^2 R_1 R_2 \eta(L_S)}\right]^{1/2}, \quad (18)$$

where the $\sigma_{Lr}$ is calculated according to either (10), (13) or (16), $\sigma_{nPI}$ is calculated according to (17), these equations aren't substituted in order to avoid excessive unwieldiness.

It must be noted, that even though the amplitude of the peak, representing the parasitic interference component is relatively small, it is wide enough and in order to find the real amplitude of the parasitic interference component $S_{PI}$, one needs to integrate the peak. As a result, for a given example $S_{PI}\simeq 10^{-5}$mW.

## 4. Experimental noise compensation

In order to support the theoretical results, an experimental study of EFPI displacement sensor resolution was carried out. Spectral measurements were performed using the optical sensor interrogator NI PXIe 4844, utilizing a tunable laser with SMF-28 single-mode fiber output. Spectrometer parameters are the following: scanning range [1.51; 1.59] µm; wavelength jitter stdev $\sigma_{\delta\lambda}$ = 1 pm; optical power $P_0 \approx 60$ µW; scale shift stdev $\sigma_{\Delta\lambda} \approx 0.05$ pm; spectral points stepping $\Delta$ = 4 pm; number of spectral points $M$ = 20001; spectrum acquisition rate about 1 Hz, spectrum acquisition time 0.035 s.

The interferometers were formed by the ends of SMF-28 fibers, packaged with PC connectors, fixed in a standard mating sleeves. Two variants of the fibers ends reflectivities were used: both the ends of SMF-28 fiber ($R_1=R_2$=3.5%); standard feeding SMF-28 fiber and an opaque metallic mirror, glued to the fiber end ($R_1$=3.5%, $R_2$≈90%). The air gaps $L_S$ and $L_R$ between the fiber ends was varied by the use of Standa 7TF2 translation stages. The reference interferometer was placed in a thermally isolated chamber in order to suppress the influence of the ambient thermal fluctuations. The experimental setup is schematically shown in fig. 6.

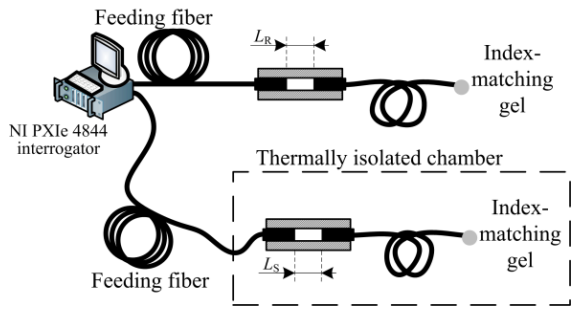

Fig. 6. Experimental setup.

### A. Matched interferometers

The aim of the first performed experiment was to verify the correlation properties of the OPD fluctuations, predicted by means of numeric modeling. As in the modeling, the OPDs of the both interferometers were set to ~200 µm, the reference interferometer OPD was fixed, while the signal interferometer OPD was scanned with step ~100 nm. In fig. 2 the experimental relation $C_{BF}(\delta L)$ (dots) is compared with the simulated dependency $C_{BF}(\delta L)$ (solid line). Not exact correspondence of simulated and experimental dependencies can be due to some discrepancy of experimental setup parameters and properties and those implied in the simulation. Also, a lower level of correlation was observed in the experiment, limiting the admissible OPD discrepancy to somewhat about 10 nm.

After that the noise compensation possibilities were tested. The OPDs of the both interferometers were set from ~100 µm up to ~800 µm with steps ~100 µm nearly equal with accuracy better than 1 nm. At each step of the OPD values the measurements of the interferometers' spectral functions were performed by two optical channels of NI PXIe 4844 interrogator during about 5 minutes, resulting in 300 measured points, with respect to them the STDs was calculated. Then the OPDs were changed and measurements were repeated for another $L_S$ and $L_R$ values and for both interferometers configurations. The dependencies of STDs of single sensor readings $L_{rS}$ and noise-suppressed double sensor readings $L_{rSC}$, calculated according to (11) on the OPD value are shown in fig. 7 (a) for the configuration with equal mirrors $R_1=R_2=3.5\%$ and in fig. 7 (b) for the configuration with $R_1=3.5\%$, $R_2\approx90\%$. Estimations of the measured OPDs' standard deviation for the setup parameters corresponding to the experimental, calculated according to (10) and (13) are also presented as thick solid and thin dashed curves for reference.

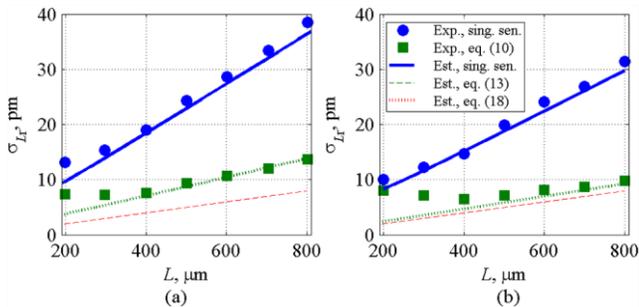

Fig. 7. Dependencies of measured OPDs fluctuations' STDs on the mean OPD value in cases with and without compensation for interferometers with $R_1=R_2=3.5\%$ (a); $R_1=3.5\%$, $R_2\approx90\%$ (b). Experimentally measured STDs are shown by dots (round – without compensation, square – with compensation), corresponding analytical estimations according to eqs. (10), (13) and (18) are shown by solid, dashed and dotted lines, respectively.

As can be seen from fig. 7, estimation according of measured OPDs' standard deviation to eq. (13) isn't in good accordance with the experimental $\sigma_{Lr}$ values. However, taking into account additional noises produced by the wavelength points jitter and parasitic interference component (see section 3.D), substituting $S_{PI}\approx10^{-5}$mW, $L_{PI}=3.2\cdot10^4$µm and eq. (13) into eq. (18), one attains the dependencies $\sigma_{Lr}(L)$, shown in fig. 7 as thick dotted curve. As a result, much better correspondence of experimental and theoretical results is obtained.

As can be seen from the fig. 7, for interferometers with $R_1=R_2=3.5\%$, the compensation effort was from ~1.5 to 2.8 with values $\sigma_{Lr}$ from 7 to 13.5 pm, whilst for scheme with $R_1=3.5\%$, $R_2\approx90\%$ the compensation effort comprised from ~1.2 to 3 with values $\sigma_{Lr}$ from 6.5 to 9.8 pm, demonstrating more uniform dependency on interferometers' OPDs.

Noticeable discrepancy between the experimental results and theoretical predictions in range of relatively small OPDs in case of noise compensation can be due to non-uniform spectrum of the wavelength points jitter with increase at lower frequencies.

### B. Non-matched interferometers

In a separate experiment, the possibility to improve the sensor resolution by means of a setup with non-matched interferometers was tested. The experimental setup was the same as the one in the previous section, except for the OPDs values: $L_S$ was set to ~314 µm and $L_R$ – to ~329 µm, with the resulting difference $\delta L$ close to the $\Delta L$, which, for the used experimental setup was 15 µm. Interferometers with mirrors $R_1=3.5\%$, $R_2\approx90\%$ were used in order to decrease the noises influence on $\sigma_{Lr}$ comparing to the scale shift influence. The signal interferometer's OPD $L_S$ was stepwise changed in the range from ~313 µm to ~315 µm in order to test the relation of compensation efficiency versus the interferometers' OPDs difference.

For each $L_S$ value, the measurements were performed for about 5 minutes, resulting in nearly 300 registered spectra of each interferometer. After that the estimated OPDs values $L_{rS}$ and $L_{rR}$ were calculated according to the approach presented in [3], whereat the noise-suppressed OPD values $L_{rSC}$ were calculated according to (15). In the fig. 8 the dependencies of the measured OPDs standard deviations with and without compensation versus the OPDs difference $\delta L$ are shown.

As can be seen from the fig. 8, the maximal attained improvement of measured OPD fluctuations level comprised twofold, whereas improvement in 1.5 times was attained for signal interferometer OPD varying in a range of more than 1 µm.

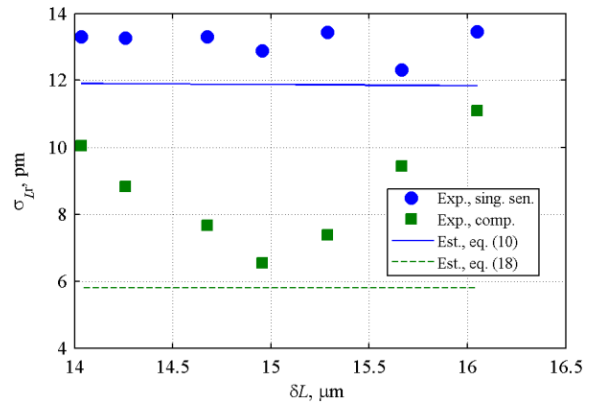

Fig. 8. Comparison of measured OPD fluctuations in case of single interferometer and noise compensation for different OPDs differences. Experimentally measured STDs are shown by dots (round – without compensation, square – with compensation), corresponding analytical estimations according to the eqs. (10) and (16) are shown by solid and dotted lines, respectively.

As can be seen in Fig. 8, fluctuations of the signal interferometer's measured OPD exceed the estimations made according to Eq. (16). This is due to nonzero correlation between interferometers' spectral functions when the difference of OPDs $\delta L$ is different from $\Delta L\sim15$µm.

## 5. Conclusion

In the current paper an original approach for eliminating the influence of the interrogating unit noises on the measured value of interferometer OPD is proposed. Parasitic OPD fluctuations, induced by unideal interrogator operation are estimated from the readings of an additional reference interferometer, which is isolated from the environmental perturbations. After that, the fluctuations of the reference interferometer's measured OPD are subtracted from the measured OPD of the signal interferometer. As a result, the STD of the signal interferometer's measured OPD values is decreased more than twofold.

The proposed technique can be utilized in order to improve the resolution of EFPI sensors in case when relatively large value of its OPD ($L \gtrsim 200$ μm) is necessary [8,21,22]. Another attractive application of the proposed noise compensation approach is resolution improvement in spectral interferometric sensing systems, utilizing relatively chip and simple interrogating units with poor wavelength scale repeatability.

## Appendix A. Relation of correlation coefficient and the compensation efficiency

In this appendix the following situation will be considered: let us have two discrete signals $x_i$ and $y_i$, each of them can be represented as a superposition of a common signal $a_i$ and white Gaussian noises $n_{xi}$ and $n_{yi}$: $x_i = a_i + n_{xi}$, $y_i = a_i + n_{yi}$. The variances of the noises are the same and equal to $\sigma^2_n$, $a_i$ variance is $\sigma^2_a = m^2 \cdot \sigma^2_n$. For simplicity, we'll assume all signals $a_i$, $n_{xi}$ and $n_{yi}$ to be zero-mean. The correlation coefficient of signals $x_i$ and $y_i$ can be written as

$$C = \frac{\sum x_i \cdot y_i}{\sqrt{\sum x_i^2 \cdot \sum y_i^2}} = \frac{\sum (a_i + n_{xi}) \cdot (a_i + n_{yi})}{\sqrt{\sum (a_i + n_{xi})^2 \cdot \sum (a_i + n_{yi})^2}} = \frac{\sum a_i^2 + \sum a_i (n_{xi} + n_{yi}) + \sum n_{xi} \cdot n_{yi}}{\sqrt{\sum (a_i^2 + 2a_i n_{xi} + n_{xi}^2) \cdot \sum (a_i^2 + 2a_i n_{yi} + n_{yi}^2)}}, \quad \text{(A1)}$$

due to statistical independence of signals $a_i$, $n_{xi}$ and $n_{yi}$, summations over the second and the third terms in numerator and the terms of form $2 \cdot a_i \cdot n_{x,y\,i}$ in denominator will result in zero. Therefore, (A1) will be written as

$$C = \frac{\sum a_i^2}{\sqrt{\left(\sum a_i^2 + \sum n_{xi}^2\right) \cdot \left(\sum a_i^2 + \sum n_{yi}^2\right)}} = \frac{\sigma_a^2}{\left(\sigma_a^2 + \sigma_n^2\right)} = \frac{m^2}{m^2 + 1}. \quad \text{(A2)}$$

Further, let us find the ratio of STDs of signals $x_i$ or $y_i$ (which are equal) and standard deviation of their difference $x_i - y_i$:

$$E = \frac{\sigma\{x\}}{\sigma\{x - y\}} = \frac{\sigma\{a + n_x\}}{\sigma\{n_x - n_y\}} = \frac{\sqrt{m^2 + 1}}{\sqrt{2}}, \quad \text{(A3)}$$

expressing $m$ from (A2) and substituting it into (A3), we obtain:

$$E = \frac{1}{\sqrt{2}} \sqrt{\frac{1}{1 - C}}. \quad \text{(A4)}$$

In context of the noise compensation, $x_i$ and $y_i$ have a sense of fluctuations of the measured OPDs $L_{rR}$ and $L_{rS}$, induced by the interrogator non-ideal operation; $a_i$ – the component of these fluctuations, induced by the common mechanisms (laser intensity noise, wavelength scale shift and points jitter); $n_{xi}$ and $n_{yi}$ – the fluctuations of $L_{rR}$ and $L_{rS}$, induced by independent noise sources (photodetector noises, uncorrelated components of wavelength scale shift and points jitter); $E$ – the compensation effort, introduced in eq. (12).

## Funding Information

The work was supported by the program State Tasks for Higher Educational Institutions (projects no. 3.1446.2014K and 2014/184).

## References

1. K. Chang, Z. Fang, K. K. Chin, R. Qu, and H. Cai, Fundamentals of Optical Fiber Sensors (John Wiley & Sons, 2012).
2. W. Wang and F. Li, "Large-range liquid level sensor based on an optical fibre extrinsic Fabry–Perot interferometer," Opt. Lasers Eng. 52, 201–205 (2014).
3. N. A. Ushakov, L. B. Liokumovich, and A. Medvedev, "EFPI signal processing method providing picometer-level resolution in cavity length measurement," in Proceedings of SPIE, B. Bodermann, K. Frenner, and R. M. Silver, eds. (2013), Vol. 8789, p. 87890Y.
4. J. Sirkis and C.-C. Chang, "Multiplexed optical fiber sensors using a single Fabry-Perot resonator for phase modulation," J. Light. Technol. 14, 1653–1663 (1996).
5. V. Bhatia, K. A. Murphy, R. O. Claus, M. E. Jones, J. L. Grace, T. A. Tran, and J. A. Greene, "Optical fibre based absolute extrinsic Fabry-Perot interferometric sensing system," Meas. Sci. Technol. 7, 58–61 (1996).
6. M. Han, Y. Zhang, F. Shen, G. R. Pickrell, and A. Wang, "Signal-processing algorithm for white-light optical fiber extrinsic Fabry-Perot interferometric sensors," Opt. Lett. 29, 1736–1738 (2004).
7. F. Shen and A. Wang, "Frequency-estimation-based signal-processing algorithm for white-light optical fiber Fabry-Perot interferometers," Appl. Opt. 44, 5206–5214 (2005).
8. Z. Wang and Y. Jiang, "Wavenumber scanning-based Fourier transform white-light interferometry," Appl. Opt. 51, 5512–5516 (2012).
9. X. Zhou and Q. Yu, "Wide-Range Displacement Sensor Based on Fiber-Optic Fabry–Perot Interferometer for Subnanometer Measurement," IEEE Sens. J. 11, 1602–1606 (2011).
10. N. A. Ushakov and L. B. Liokumovich, "Resolution limits of extrinsic Fabry–Perot interferometric displacement sensors utilizing wavelength scanning interrogation," Appl. Opt. 53, 5092–5099 (2014).
11. D. Tosi, S. Poeggel, G. Leen, and E. Lewis, "Adaptive filter-based interrogation of high-sensitivity fiber optic Fabry-Perot interferometry sensors," Sensors Actuators A Phys. 206, 144–150 (2014).
12. C. Ma and A. Wang, "Signal processing of white-light interferometric low-finesse fiber-optic Fabry-Perot sensors," Appl. Opt. 52, 127–138 (2013).
13. Y. Jiang, "Fourier Transform White-Light Interferometry for the Measurement of Fiber-Optic Extrinsic Fabry – Pérot Interferometric Sensors," IEEE Photonics Technol. Lett. 20, 75–77 (2008).
14. Z. Yu and A. Wang, "Fast White Light Interferometry Demodulation Algorithm for Low-Finesse Fabry – Pérot Sensors," IEEE Photonics Technol. Lett. 27, 817–820 (2015).
15. J. Hough, "The stabilisation of lasers for interferometric gravitational wave detectors," in The Detection of Gravitational Waves (1991), pp. 329–351.
16. Y. Jiang and C. Tang, "A high-resolution technique for strain measurement using an extrinsic Fabry–Perot interferometer (EFPI) and a compensating EFPI," Meas. Sci. Technol. 19, 065304 (2008).
17. N. Ushakov and L. Liokumovich, "Measurement of dynamic interferometer baseline perturbations by means of wavelength-scanning interferometry," Opt. Eng. 53, 114103 (2014).
18. D. C. Rife and R. R. Boorstyn, "Single-Tone Parameter Estimation from Discrete-Time Observations," IEEE Trans. Inf. Theory 20, 591–598 (1974).
19. H. Fu and P. Y. Kam, "MAP / ML Estimation of the Frequency and Phase of a Single Sinusoid in Noise," IEEE Trans. Signal Process. 55, 834–845 (2007).
20. N. A. Ushakov and L. B. Liokumovich, "Multiplexed Extrinsic Fiber Fabry-Perot Interferometric Sensors: Resolution Limits," J. Light. Technol. 33, 1683–1690 (2015).
21. E. A. Moro, M. D. Todd, and A. D. Puckett, "Dynamics of a noncontacting, white light Fabry-Perot interferometric displacement sensor," Appl. Opt. 51, 4394–402 (2012).

22. Y. Shen, Z. Ding, Y. Yan, C. Wang, Y. Yang, and Y. Zhang, "Extended range phase-sensitive swept source interferometer for real-time dimensional metrology," Opt. Commun. 318, 88–94 (2014).